\newif\ifdoubleblind
\newif\ifacm
	
\ifacm
	\documentclass[sigconf]{acmart}
\else
	\documentclass[conference]{IEEEtran}
\fi

\usepackage{amsmath}

\usepackage{multirow}
\usepackage{tabularx,booktabs}
\usepackage{xspace}

\usepackage[binary-units=true]{siunitx}
\usepackage{tikz}
\usetikzlibrary{positioning}

\ifacm
	\usepackage[nolist]{acronym}
\usepackage{booktabs} 
\usepackage{subfig}
\usepackage{balance}

\usepackage{array}
\usepackage{graphicx}
\usepackage{wasysym}
\usepackage{xparse}

\usepackage{makecell}
\newcolumntype{Y}{>{\centering\arraybackslash}X}

\hypersetup{draft}

\renewcommand\footnotetextcopyrightpermission[1]{} 
\else
	\usepackage{graphicx}
\usepackage{epstopdf}
\usepackage{standalone}
\usepackage[nolist ]{acronym}
\usepackage{tcolorbox}
\usepackage{pdfpages}
\usepackage[bookmarks=false]{hyperref}
\usepackage{pdfcomment}
\usepackage{hologo}

\usepackage{xstring}

\usepackage[utf8]{inputenc}
\usepackage{marvosym}

\usepackage{algorithm}
\usepackage{algpseudocode}

\usepackage{psfrag}
\usepackage{graphicx}

\usepackage{rotating}

\renewcommand{\baselinestretch}{1}

\usepackage[caption=false,font=footnotesize]{subfig}
\usepackage{stfloats}
\fi

\usepackage{makecell}

\newcommand{\SG}{Samsung Galaxy S21 \ac{5G}}

\usepackage{color, colortbl} 
\usepackage{amssymb}
\usepackage{balance}

\usepackage{import}

\hypersetup{draft}

\begin{document}

\newcommand{\paperTitle}{System Modeling and Performance Evaluation of \\Predictive QoS for Future Tele-Operated Driving}
\newcommand{\paperAuthors}{Hendrik Schippers, Cedrik Schüler, Benjamin Sliwa, and Christian Wietfeld}
\newcommand{\paperEmails}{$\{$Hendrik.Schippers, Cedrik.Schueler, Benjamin.Sliwa, Christian.Wietfeld$\}$@tu-dortmund.de}

\newcommand\single{1\textwidth}
\newcommand\double{.48\textwidth}
\newcommand\triple{.32\textwidth}
\newcommand\quarter{.24\textwidth}
\newcommand\singleC{1\columnwidth}
\newcommand\doubleC{.475\columnwidth}

\newcommand{\figurePadding}{0pt}
\newcommand{\figureTopPadding}{\figurePadding}
\newcommand{\figureBottomPadding}{\figurePadding}

\newcommand\tikzFig[2]
{
	\begin{tikzpicture}
		\node[draw,minimum height=#2,minimum width=\columnwidth,text width=\columnwidth,pos=0.5]{\LARGE #1};
	\end{tikzpicture}
}

\newcommand{\dummy}[3]
{
	\begin{figure}[b!]  
		\begin{tikzpicture}
		\node[draw,minimum height=6cm,minimum width=\columnwidth,text width=\columnwidth,pos=0.5]{\LARGE #1};
		\end{tikzpicture}
		\caption{#2}
		\label{#3}
	\end{figure}
}

\newcommand\pos{h!tb}

\newcommand{\basicFig}[7]
{
	\begin{figure}[#1]  	
		\vspace{#6}
		\centering		  
		\includegraphics[width=#7\columnwidth]{#2}
		\caption{#3}
		\label{#4}
		\vspace{#5}	
	\end{figure}
}
\newcommand{\fig}[4]{\basicFig{#1}{#2}{#3}{#4}{0cm}{0cm}{1}}

\newcommand\sFig[2]{\begin{subfigure}{#2}\includegraphics[width=\textwidth]{#1}\caption{}\end{subfigure}}
\newcommand\vs{\vspace{-0.3cm}}
\newcommand\vsF{\vspace{-0.4cm}}

\newcommand{\subfig}[3]
{%
	\subfloat[#3]%
	{%
		\includegraphics[width=#2\textwidth]{#1}%
	}%
	\hfill%
}

\newcommand\circled[1] 
{
	\tikz[baseline=(char.base)]
	{
		\node[shape=circle,draw,inner sep=1pt] (char) {#1};
	}\xspace
}
\begin{acronym}
	\acro{xyzabc}{asdasd}
\acro{4G}{4th Generation of Mobile Communication Networks}
\acro{5G}{5th Generation of Mobile Communication Networks}
\acro{6G}{6th Generation of Mobile Communication Networks}

\acro{5GAA}{\ac{5G} Automotive Association} 

\acro{ECDF}{Empirical Cumulative Distribution Function}

\acro{HiL}{Hardware-in-the-loop}

\acro{KPI}{Key Performance Indicator}

\acro{LIDAR}{Light Detection and Ranging}
\acro{ML}{Machine Learning}
\acro{mmWave}{Millimeter-Wave}
\acro{MNO}{Mobile Network Operator}

\acro{NR}{New Radio}
\acro{NSA}{Non-Standalone}
\acro{QoE}{Quality of Experience}
\acro{QoS}{Quality of Service}

\acro{REM}{Radio Environmental Map}
\acro{RMSE}{Root Mean Squared Error}

\acro{RAT}{Radio Access Technology}
\acro{RF}{Random Forest}
\acro{RIS}{Reconfigurable Intelligent Surface}
\acro{RS}{Reference Signal}
\acrodefplural{RS}{Reference Signals}
\acro{RSRP}{Reference Signal Received Power}
\acro{RSRQ}{Reference Signal Received Quality}
\acro{RSSI}{Received Signal Strength Indicator}
\acro{RTT}{Round-Trip Time}

\acro{SINR}{Signal to Interference and Noise Ratio}
\acro{SS-RSRP}{Syncronization Signal \ac{RSRP}}
\acro{TCP}{Transmission Control Protocol}
\acro{ToD}{Tele-operated Driving}

\acro{UDP}{User Datagram Protocol}
\acro{UE}{User Equipment}

\acro{V2X}{Vehicle-to-Everything }
\end{acronym}

\title{\paperTitle}

\ifacm
\else

	\title{\paperTitle}

	\ifdoubleblind
	\author{\IEEEauthorblockN{\textbf{Anonymous Authors}}
		\IEEEauthorblockA{Anonymous Institutions\\
			e-mail: Anonymous Emails}}
	\else
	\author{\IEEEauthorblockN{\textbf{\paperAuthors}}
		\IEEEauthorblockA{Communication Networks Institute,	TU Dortmund University, 44227 Dortmund, Germany\\
			e-mail: \paperEmails}}
	\fi
	
	\maketitle

\fi




\begin{abstract}\acused{5G}\acused{4G}\acused{6G}
Future \ac{ToD} applications place challenging \ac{QoS} demands on existing mobile communication networks that are of highly important to comply with for safe operation. New remote control and platooning services will emerge and pose high data rate and latency requirements. One key enabler for these applications is the newly available \ac{5G} \ac{NR} promising higher bandwidth and lower latency than its predecessors. In addition to that, public \ac{5G} networks do not consistently deliver and do not guarantee the required data rates and latency of \ac{ToD}.\\
In this paper, we discuss the communication-related requirements of tele-operated driving. \ac{ToD} is regarded as a complex system consisting of multiple research areas. One key aspect of \ac{ToD} is the provision and maintenance of the required data rate for teleoperation by the mobile network. An in-advance prediction method of the end-to-end data rate based on so-called \acp{REM} is discussed. Furthermore, a novel approach improving the prediction accuracy is introduced and it features individually optimized \ac{REM} layers. \\
Finally, we analyze the implementation of tele-operated driving applications on a scaled vehicular platform combined with a cyber-physical test environment consisting of real and virtual objects. This approach enables large-scale testing of remote operation and autonomous applications.

\end{abstract}

\begin{tikzpicture}[remember picture, overlay]
\node[below=5mm of current page.north, text width=20cm,font=\sffamily\footnotesize,align=center] {Accepted for presentation in: 2022 Annual IEEE International Systems Conference (SysCon)\vspace{0.3cm}\\\pdfcomment[color=yellow,icon=Note]{
@InProceedings\{Schippers2022a,\\
	Author = \{Hendrik Schippers and Cedrik Schüler and Benjamin Sliwa and Christian Wietfeld\},\\
	Title = \{System Modeling and Performance Evaluation of Predictive QoS for Future Tele-Operated Driving\},\\
	Booktitle = \{2022 Annual IEEE International Systems Conference (SysCon)\},\\
	Year = \{2022\},\\
	Address = \{Montreal, Canada\},\\
	Month = \{apr\},\\
\}
}};
\node[above=5mm of current page.south, text width=15cm,font=\sffamily\footnotesize] {2022~IEEE. Personal use of this material is permitted. Permission from IEEE must be obtained for all other uses, including reprinting/republishing this material for advertising or promotional purposes, collecting new collected works for resale or redistribution to servers or lists, or reuse of any copyrighted component of this work in other works.};
\end{tikzpicture}

\maketitle

\section{Introduction}

%
%

\begin{figure}
	\centering
	\includegraphics[width=\columnwidth]{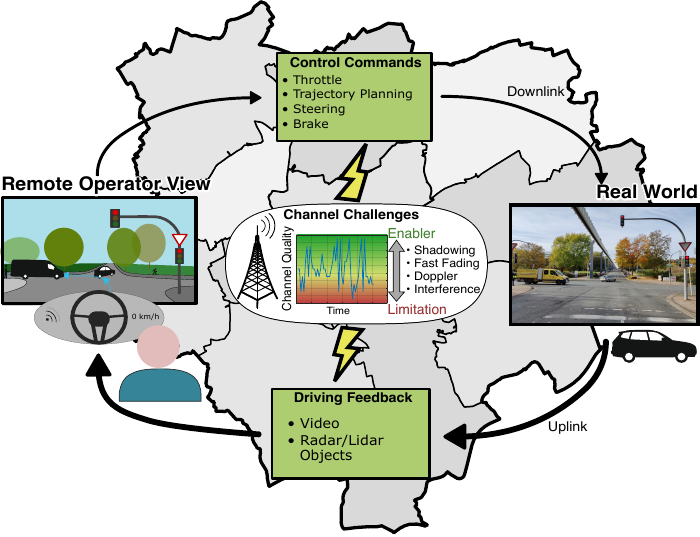}
	\caption{System overview of tele-operated driving applications in the context of mobile networking. (Map: $\copyright$ OpenStreetMap contributors, CC BY-SA)}
	\label{fig:systemOverview}
\end{figure}

\acl{ToD} (\acs{ToD}) can be seen as the next step towards automated driving and might even prevent possible driver shortages in industries like truck-based logistics as seen recently in Great Britain. Teleoperators might aid drivers or autonomous vehicles in challenging situations and \ac{ToD} could be coupled with platooning. Several challenges concerning the sensors and actuators in the vehicle, but also regarding the mobile network enabling the teleoperation, need to be addressed. \ac{ToD} presumes high data rates and low latency in addition to other requirements \cite{AutomotiveAssociation/2019a,AutomotiveAssociation/2020a}. That is why some kind of predictive \ac{QoS} or even network quality guarantees are needed for teleoperation to work correctly \cite{AutomotiveAssociation/2019a}.\\
Based on the current channel conditions, the mobile network channel can either be an enabler or a limitation, as demonstrated in \autoref{fig:systemOverview}. Due to mobility-dependent impact factors such as shadowing, fast fading and interference from other users, the \ac{QoS} can be severely diminished. It is of crucial importance to predict this limitation of the channel to preserve the \ac{QoE} of the user remotely steering a vehicle \cite{AutomotiveAssociation/2020a}. Based on the information of video or object data transmitted via the mobile network, the operator can directly control the vehicle or set trajectories, which the vehicle independently follows. These commands must be executed reliably and immediately.\\

%
%
The contributions of this paper are summarized as follows:
%
%
\begin{itemize}
	\item Derivation of a \textbf{system model} and development of an evaluation approach for \ac{ToD} and autonomous driving.
	\item Elaborate key \textbf{performance requirements} for \ac{ToD} and performing a \ac{5G} \ac{NSA} \textbf{measurement campaign} as a base for further investigation.
	\item Utilization of \textbf{multi-dimensional \acp{REM}} to proactively enable end-to-end predictive \ac{QoS}.
\end{itemize}

%
%
The remainder of the paper is structured as follows. After discussing the related work in \autoref{sec:related_work}, the system modeling of \ac{ToD} and methodological aspects from a predictive-\ac{QoS} perspective are shown in \autoref{sec:approach}. In \autoref{sec:datarate}, detailed results concerning end-to-end data rate predictions via \acp{REM} are provided. Finally, an approach to systematically evaluate autonomous driving in a cyber-physical area is discussed.

\section{Related Work} \label{sec:related_work}

Many previous research works have recognized the potentials of \ac{ToD} for future intelligent traffic systems. However, there is no ready-proven \ac{ToD} application available on a large scale or even outside of a dedicated test environment.\\
One major challenge are the high data rate and latency requirements that need to be met. We summarized system requirements of several related works in \autoref{tab:parameters}. Different related works assume varying velocities and working modes, as there is no standardization for \ac{ToD}. That results in differing requirements and complicates a comparison. Furthermore, there is progress within requirement statements, as the \ac{5GAA} has changed its thresholds from 2020 \cite{AutomotiveAssociation/2019a} to 2021 \cite{AutomotiveAssociation/2021a}.\\
However, all sources are consistent that a high service level requirement of over \SI{99}{\percent} is mandatory for \ac{ToD}.  Also, all sources agree that a higher data rate in the uplink is needed than in the downlink because video and object data need to be transmitted in the uplink direction. Demand-based pattern usage is discussed in \cite{Bektas/etal/2021a}. Differences are found in the number of cameras and the video quality settings, which result in highly varying uplink data rate requirements. While the authors of \cite{Neumeier/etal/2019a} state an uplink data rate requirement of \SI{3}{\mega\bit\per\second}, the authors of \cite{Katranaras/etal/2021a} and \cite{Schimpe/etal/2020a} demand for up to \SI{50}{\mega\bit\per\second}. In the downlink, relatively small data rates lower or equal than \SI{5}{\mega\bit\per\second} are required. 

\newcommand{\mrow}[1]{\multirow{2}{*}{#1}}
\newcommand{\requ}[8]
{
\mrow{#1} & \mrow{#2} & DL: \thead[r]{#3} & DL: \thead[r]{#5} & DL: \thead[r]{#7} \\
& & & UL: \thead[r]{#4} & UL: \thead[r]{#6} & UL: \thead[r]{#8} \\
}

\newcommand{\myref}[2]{#1 \cite{#2}}

\newcommand*\fullcirc[1][1ex]{\tikz\fill (0,0) circle (#1);} 
\definecolor{Gray}{gray}{0.9}
\definecolor{White}{gray}{1}

\begin{table}[ht]
	\centering
	\caption{Tele-operated driving minimum network requirements}
	\begin{tabular}{lrrrrrrr}
		\toprule
		 \multirow{2}{*}{\rotatebox{90}{\textbf{Source}}}& \thead[c]{\textbf{$V_{Max}$}\\ \textbf{[km/h]}}&\multicolumn{2}{c}{\thead[c]{\textbf{Data Rate} \\\textbf{[Mbit/s]}}} & \multicolumn{2}{c}{\thead[c]{\textbf{Service Level}\\\textbf{Reliability [\%]}}} & \multicolumn{2}{c}{\thead[c]{\textbf{Latency}\\\textbf{[ms]}}}\\
	&&DL&UL&DL&UL&DL&UL\\

		\midrule
		\multicolumn{8}{c}{\emph{Included in 5GAA System requirements analysis and architecture \cite{AutomotiveAssociation/2021a}}} \\ 
		\midrule
				
		 \cite{AutomotiveAssociation/2021b} &50&0.4&32/36&{\footnotesize99.999}&99&20&100\\ 
		 \cite{Katranaras/etal/2021a}&--&5&8--50&{\footnotesize99.999}&99&{\footnotesize 10--66}&{\footnotesize 10--50}\\
		 \midrule
		 \cite{Schimpe/etal/2020a},\cite{Floess/etal/2020a}&15&0.5&10--50&99.9&99&--&40\\
		 \cite{5G-MOBIX}&8&--&--&--&--&80&120\\ 
		 \midrule
		 \rowcolor{Gray}
		 \cite{AutomotiveAssociation/2021a}&15&0.3&8--30&99&99.9&\multicolumn{2}{c}{300}\\
		\midrule
		\multicolumn{8}{c}{\emph{Other references}} \\ 
		\midrule
		\cite{3GPP/2020a}&250&1&25&\multicolumn{2}{c}{99.999}&5&5\\
		\cite{Neumeier/etal/2019a}&- & 0.25& 3& -&-&\multicolumn{2}{r}{250} \\

%
%
%
%
		\bottomrule
		\\
	\multicolumn{8}{l}{$\square$ : Direct Control Tele-operated Driving}\\
	\multicolumn{8}{l}{\textcolor{Gray}{$\blacksquare$} : Indirect Control Tele-operated Driving}\\
	\end{tabular}
	\label{tab:parameters}
\end{table}

There are two main modes of tele-operated driving: \textit{indirect} and \textit{direct control} \ac{ToD} \cite{AutomotiveAssociation/2021a}.\\
With direct control, an operator directly controls the steering wheel, the accelerator pedal and other actuators. The vehicle only has to be able to sustain a lower level of automation to securely come to a halt if the connection to the teleoperator is cut.  In the case of indirect \ac{ToD}, the vehicle has to reach a significantly higher level of automated driving \cite{AutomotiveAssociation/2021a} because the operator can set trajectories the vehicle needs to follow. It is striking, that in this case, a control latency of only \SI{300}{\milli\second} needs to be reached compared to a latency of roughly \SI{100}{\milli\second} in the case of direct control \cite{AutomotiveAssociation/2021a}. Thus, some requirements for the network \acp{KPI} are reduced.\\

Despite several sources specifying uplink and downlink latency requirements, the \ac{5GAA} states that only the whole \ac{RTT} is the main determining factor \cite{AutomotiveAssociation/2021a}. In turn, the maximum allowed \ac{RTT} depends on the maximum velocity allowed to operate the vehicle at \cite{AutomotiveAssociation/2021a}: Higher velocities demand even faster reaction times for braking and evasive maneuvers.\\
With a higher latency, precise driving operations get increasingly difficult \cite{Neumeier/etal/2019a}. That is a major challenge for the mobile network in the case of direct control \ac{ToD}. For indirect \ac{ToD}, the latency is less critical, but the high data rate that needs to be fulfilled continuously for operation is still concerning.\\
Sudden service interruptions during \ac{ToD} could be fatal and thus need to be prevented. That could be done with the help of predictive \ac{QoS}. Various related works have performed data rate predictions of mobile networks. These can either be performed instantaneous like in \cite{Palaios/etal/2021a} and \cite{Sliwa/Wietfeld/2019a} or with the help of previously generated so-called \acp{REM} \cite{Sliwa/etal/2018a,Sliwa/etal/2021a}.\\
Most of these predictions are based on a measurement study contributing a training set of feature vectors and labels (achieved data rate). The features are commonly a subset of real-time measured passive \acp{RS}. Machine learning is used to generalize this data set to predict the achieved data rate based on new feature vectors. Many related works use tree-based regression models like \acp{RF} for their predictions \cite{Raca/etal/2017a,Samba/etal/2017a,Jomrich/etal/2018a}. \\
This methodology is promising to enable in advance data rate predictions for \ac{ToD} with the goal to reduce service interruptions and improve the \ac{QoE}.
\section{System modeling of predictive-QoS for future tele-operated} \label{sec:approach}

\begin{figure*}
	\centering
	\includegraphics[width=\single]{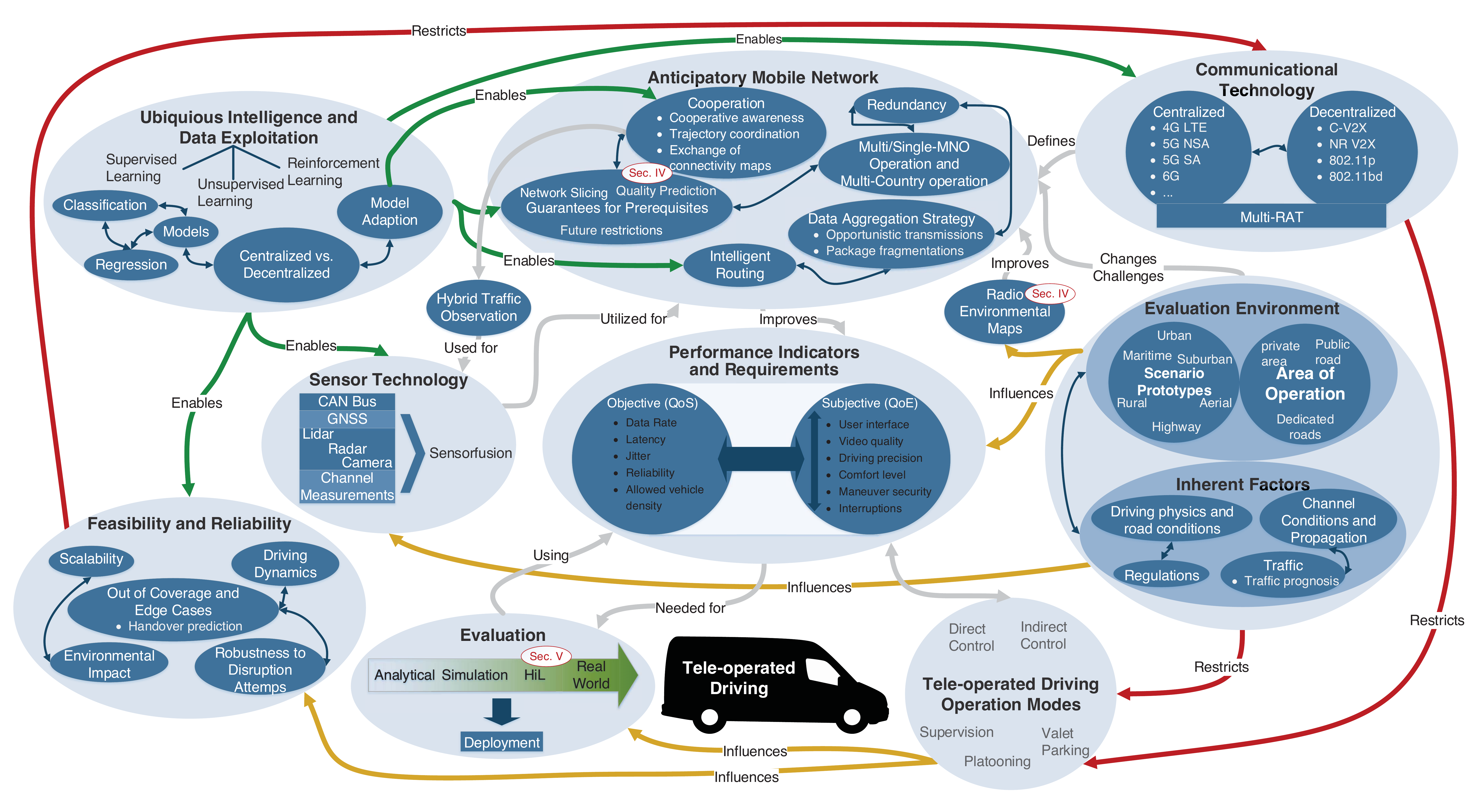}
	\caption{Systemigram of the system architecture model for data-driven optimization of mobile traffic consisting of data acquisition, data processing and data exploitation to enable new Services in varying environments.}
	\label{fig:systemigram}
\end{figure*}

Different operation modes of tele-operated driving set requirements for performance indicators to be met, as described in the previous section. These influence the feasibility parameters like the possible driving dynamics or the needed network coverage as demonstrated in \autoref{fig:systemigram}. More complex operation modes demand for more sophisticated communication technology like \ac{5G}, future \ac{6G} or even a combined approach of several communication technologies in a multi-\ac{RAT} approach. Centralized and decentralized technologies might be needed to complement their mutual strengths in an attempt to maximize coverage and user experience. While centralized approaches can reach higher data rates due to resource allocation by a central entity, decentralized solutions can also work in regions with otherwise insufficient coverage.\\
That is why the evaluation environment does also affect tele-operated driving. While huge competition for radio resources can be expected inside an urban environment, an overall inferior coverage might be the case in rural settings. Both pose challenges to the communication technology. Despite these challenges, data transfers need to be reliable and of low latency as independent of the channel conditions and the expected traffic as possible.\\

Network slicing and data rate predictions can be utilized to prevent and forecast inadequate network states. With the help of network slicing, network resources can be reserved and allocated to a specific user.  In addition to that, multi-\ac{MNO} approaches can improve the coverage significantly with the help of redundant connections \cite{Sliwa/Wietfeld/2019a}. Based on the criticality of the data, different aggregation strategies can be used to prevent unnecessary transmissions in grim coverage situations. Data aggregation can be coupled with intelligent routing algorithms, which can also enable cooperative sensor fusion between tele-operated vehicles. The more sensor data available, the more convenient and secure tele-operated driving can be. For example, trajectory coordination combined with aggregated sensor data from several vehicles can be used to warn drivers of obstacles and driving paths, which would usually be invisible to their perspective.\\

Key enablers for many of these features and applications are various kinds of machine learning.\\
\textbf{Machine learning} can be divided into three categories, which can be used to solve complex challenges across multiple research areas. Different approaches in several disciplines need to be evaluated for future tele-operated driving applications. 
\begin{itemize}
	\item \texttt{Supervised Learning} is based on previously labeled data. A machine learning model is trained on this data to find patterns and be able to generalize onto unseen data. This kind of machine learning is arguably the most common type.
	\item \texttt{Unsupervised Learning} can be used to autonomously extract hidden patterns in unlabeled data and cluster them.
	\item \texttt{Reinforcement Learning} rewards an entity based on its actions. Based on the rewards, the entity changes its strategy with the goal of getting as much reward as possible for its actions taken.
\end{itemize}
All these classes of machine learning are applied in one or more parts needed for tele-operated driving, like challenges in routing relying on reinforcement learning, supervised learning is used for predictive \ac{QoS} and unsupervised leaning can be used for trajectory planning and coordination. Not only centralized approaches but also decentralized approaches are necessary to adapt existing models appropriately.\\
Based on the type of the label, the machine learning models try to predict the process is called \texttt{Classification} or \texttt{Regression}. Classification means predicting a set of classes, whereas regression means the closest possible accurate approximation of a number.\\
Machine learning models can be trained on a central computation entity or decentralized on many devices. Data needs to be directed to the central entity to be able to centrally train a model. That may conflict with local authorities or be impossible due to the inherent data transfers generating too much traffic \cite{Park/etal/2021a}.\\
For predictive \ac{QoS} applications, supervised learning models are most commonly used in a regression configuration.\\
In this paper, the python library \texttt{scikit-learn} \cite{scikitlearn} is used for machine learning.\\

With the goal of deploying a real-world capable tele-operated driving application, an evaluation process has to be gone through, starting with analytical calculations. These can be used as a base for simulations and later hardware in the loop experiments. During this process, repeated tests of compliance with performance indicators and requirements need to be done. Only then, real-world evaluations can be conducted. These consist of pre-tested individual parts of the tele-operated application that are composed into one system.\\

\textbf{Geospatial aggregation of network context information with Radio Environmental Maps (REM)}: As stated by the \ac{5GAA} \cite{AutomotiveAssociation/2021a}, predictive \ac{QoS} is a key enabler for \Ac{ToD}, as it can ensure service availability and user experiences. One method to achieve in advance network quality predictions is based on \acp{REM} \cite{Sliwa/etal/2018a,Sliwa/etal/2019a,Li/etal/2018a}.\\
\begin{figure*}
	\includegraphics[width=\linewidth]{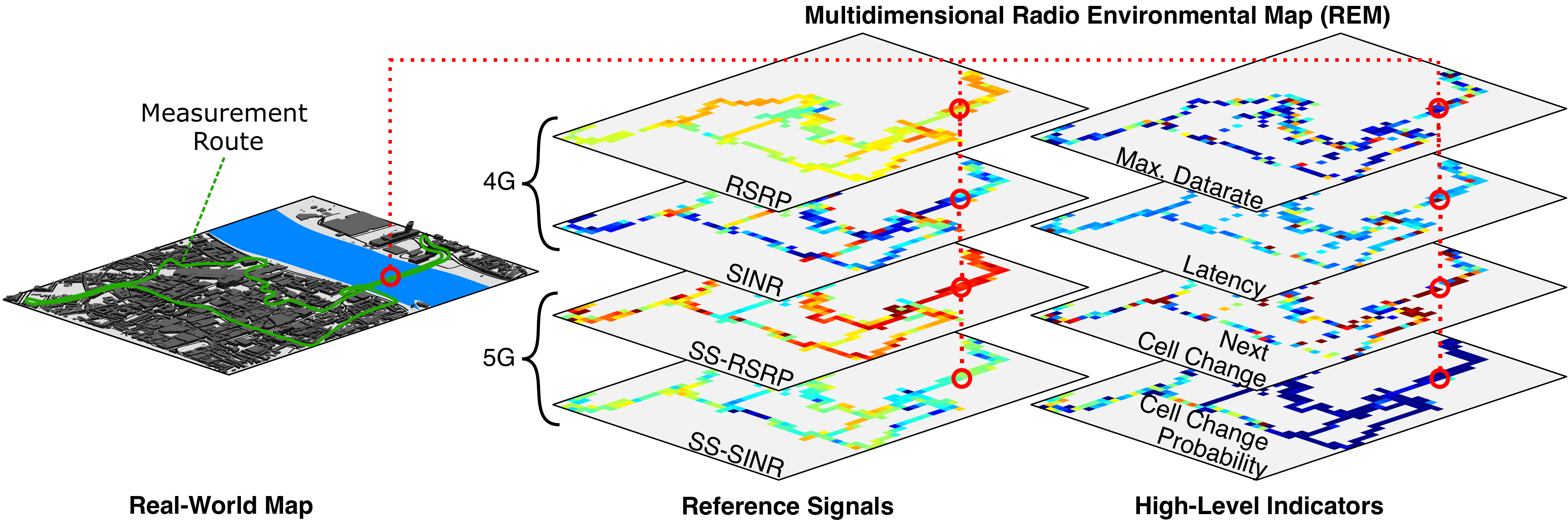}
	\caption{Real-world map and radio environmental map of the city center of Cologne with a cell width of \SI{50}{\metre} consisting of reference signals for \acl{4G} (\acs{4G}) and \ac{5G} mobile networks and combined or higher-level indicators. (Map: $\copyright$ OpenStreetMap contributors, CC BY-SA)}
	\label{fig:remExample}
\end{figure*}
\acp{REM} consist of a predefined multidimensional grid containing aggregated reference signals or performance metrics based on geographical locations \cite{Jomrich/etal/2018a}. These are commonly aggregated over several measurements of one person or a crowd-sensing approach \cite{Sliwa/etal/2020a}. However, it is also possible to calculate a \ac{REM} utilizing simulation methods such as ray-tracing. The \ac{REM} can then be used as a digital twin of the radio domain in a specific area. An example of a \ac{REM} with a cell width of \SI{50}{\metre} can be seen in \autoref{fig:remExample}. Passive \acp{RS} like the \ac{RSRP} and the \ac{SINR} are utilized.\\
In addition to passively measurable \ac{4G} and \ac{5G} \acp{RS}, high-level indicators resulting from active measurements or further processing of raw measurements are integrated. It has to be noted that additional parameters like the \ac{RSRQ} are also used inside \acp{REM}, but these are not shown due to clarity and space considerations.\\

Analogously to instantaneous measurements, network quality predictions can be based on the features located inside the \ac{REM}: To predict a data rate at a specific location, the corresponding feature vector stored in the \ac{REM} is extracted and fed into a trained machine learning model \cite{Sliwa/etal/2018a,Sliwa/etal/2020a}. \\
The quality prediction based on \acp{REM} has further advantages in addition to the ability to predict future data rates. Due to the accumulation of measurements, over-fitting to specific radio channel situations can be prevented \cite{Sliwa/etal/2020a}. With the help of predicted data rates, the energy consumption and the spectrum usage can be improved \cite{Li/etal/2018a} if the driven trajectory is known or can be predicted, and delays of the transmission can be tolerated \cite{Sliwa/etal/2018a,Sliwa/etal/2019a}. Furthermore, appropriate trajectories for \ac{ToD} can be chosen based on \ac{REM} information \cite{Kulzer/etal/2021a}.

One key factor of \acp{REM} is the cell width $c$ of the underlying grid \cite{Sliwa/etal/2020a}. Not only the spatial resolution and thus the precision and occupied memory of the \ac{REM} is affected. If the \ac{REM} is built by measurements, care must be taken to ensure the \ac{REM} is without gaps. Otherwise lookup misses occur, which need to be treated separately. As the rate of misses is dependent on the cell width, the probability of misses rise with a higher spatial resolution of the \ac{REM}. As a result, the cell width has to be adapted to minimize the prediction error \cite{Sliwa/etal/2020a}.\\

\section{Real World End-to-End Data Rate Prediction and Evaluation using Radio Environmental Maps} \label{sec:datarate}
As described in the previous section, one method to reduce the impact of fluctuating data rates of public mobile networks on the \ac{QoE} of tele-operated driving applications is \acp{REM}-aided in advance data rate prediction. Large-scale real-world measurements are needed to evaluate this setup. \\
Over 7000 data rate and 35~000 \acp{RS} measurements have been conducted in several areas in the German federal state of North Rhine-Westfalia using the \SG in a public \ac{5G} \ac{NSA} mobile network. In the \ac{NSA} mode, an existing \ac{4G} core network is used instead of a dedicated \ac{5G} core network. Major cities like Cologne, Bonn and Dortmund, highways and also suburban and urban areas are covered. A dedicated \texttt{Android} application developed for this purpose has been utilized, first used in \cite{Sliwa/etal/2021a}. The open-source software \texttt{iperf 3.9} was used internally to conduct the data rate measurements.

In this work, the measurements using the \ac{UDP} protocol in the uplink direction with a payload size of \SI{5}{\mega\byte} to \SI{10}{\mega\byte} for each individual measurement are utilized to emulate \ac{ToD} video streaming and object data transmission. The raw measurements and the used \texttt{Android} application are publicly available under this link \cite{Schippers/2021a}.\\
\begin{figure}
	\centering
	\def\svgwidth{\columnwidth}
	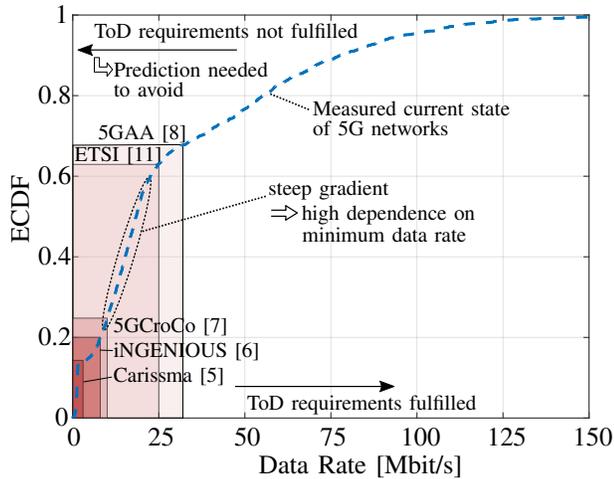
	\caption{Empirical Cumulative distribution function (ECDF) of the measured data rate with the \SG~using the \ac{UDP} protocol in the uplink direction and the minimum uplink data rate to enable tele-operated driving according to related work. The raw measurements are available at \cite{Schippers/2021a}.}
	\label{fig:dataratecdf}
\end{figure}

An \ac{ECDF} of the achieved \ac{UDP} uplink data rates is shown in \autoref{fig:dataratecdf}. On top of this, the minimum uplink data rate requirements of several related works are displayed. All data rate requirements are below \SI{50}{\mega\bit\per\second}. In \SI{75}{\percent} of all measurements, a data rate lower than \SI{50}{\mega\bit\per\second} is reached. Around and below this data rate, the gradient of the \ac{ECDF} is steep. Consequently, slightly varying data rate requirements have a significant impact on the number of places where \ac{ToD} would be feasible with the current \ac{5G} technology.
In the case of the most challenging requirements of \SI{30}{\mega\bit\per\second}, in over \SI{60}{\percent} of the measurements, \ac{ToD} would not be possible. In the case of the least challenging requirements, in over \SI{80}{\percent}, \ac{ToD} would be possible. That results in over \SI{40}{\percent} of the measurements being located in a range where it is uncertain if \ac{ToD} is possible. Independent of the proposed needed data rate, at a significant part of the measurements, \ac{ToD} would not be possible.\\

This observation further underlines the criticality of the prediction of where \ac{ToD} would not be possible to prevent frequent service interruptions.\\
However, the achieved data rate depends not only on the coverage, bandwidth and network technology at a specific location but also on resource competition with other users. Based on one indicator alone, an accurate forecast of the data rate is not possible, because multiple factors influence the achieved data rate. That is why data rate prediction is a demanding task, which needs to be based on a set of multiple indicators, including network, application and mobility context parameters.\\
Not all influencing factors like the activity of other users can be measured directly. However, the \ac{UE} can measure a set of \acp{RS} giving insight into the current channel conditions. For example, the signal quality represented by the \ac{RSRQ} can indicate a crowded radio channel and thus an expected lower data rate. 

\begin{figure}
	\centering
	\includegraphics[width=0.9\columnwidth]{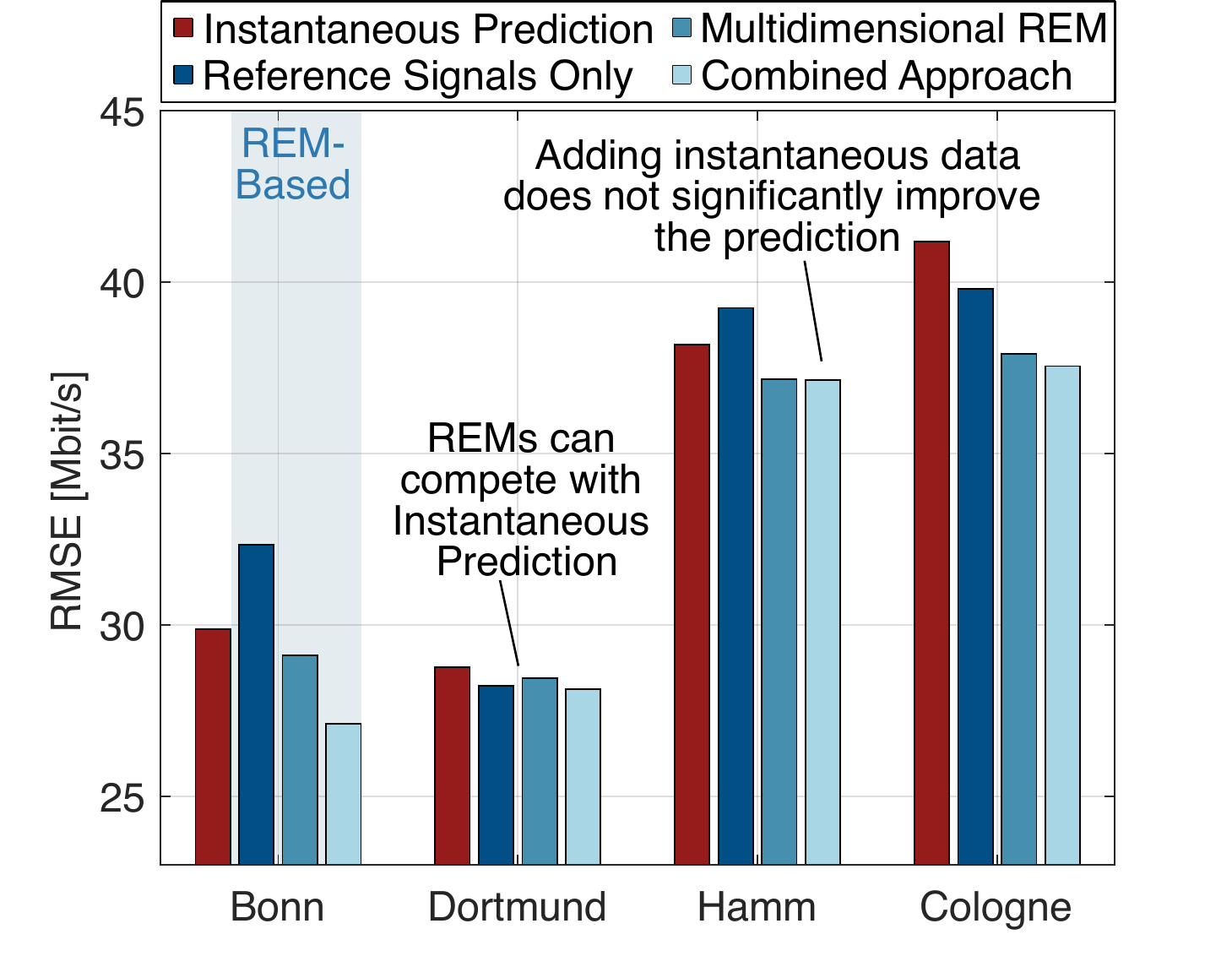}
	\caption{Data rate prediction performance comparison: Two \ac{REM} approaches with a cell size of \SI{50}{\metre}, one instantaneous prediction approach as the baseline and a hybrid approach using both the instantaneous and the \ac{REM} data, are examined. The results are evaluated on different scenarios.}
	\label{fig:datarateprediction}
\end{figure}

\begin{figure}
	\centering
	\includegraphics[width=\linewidth]{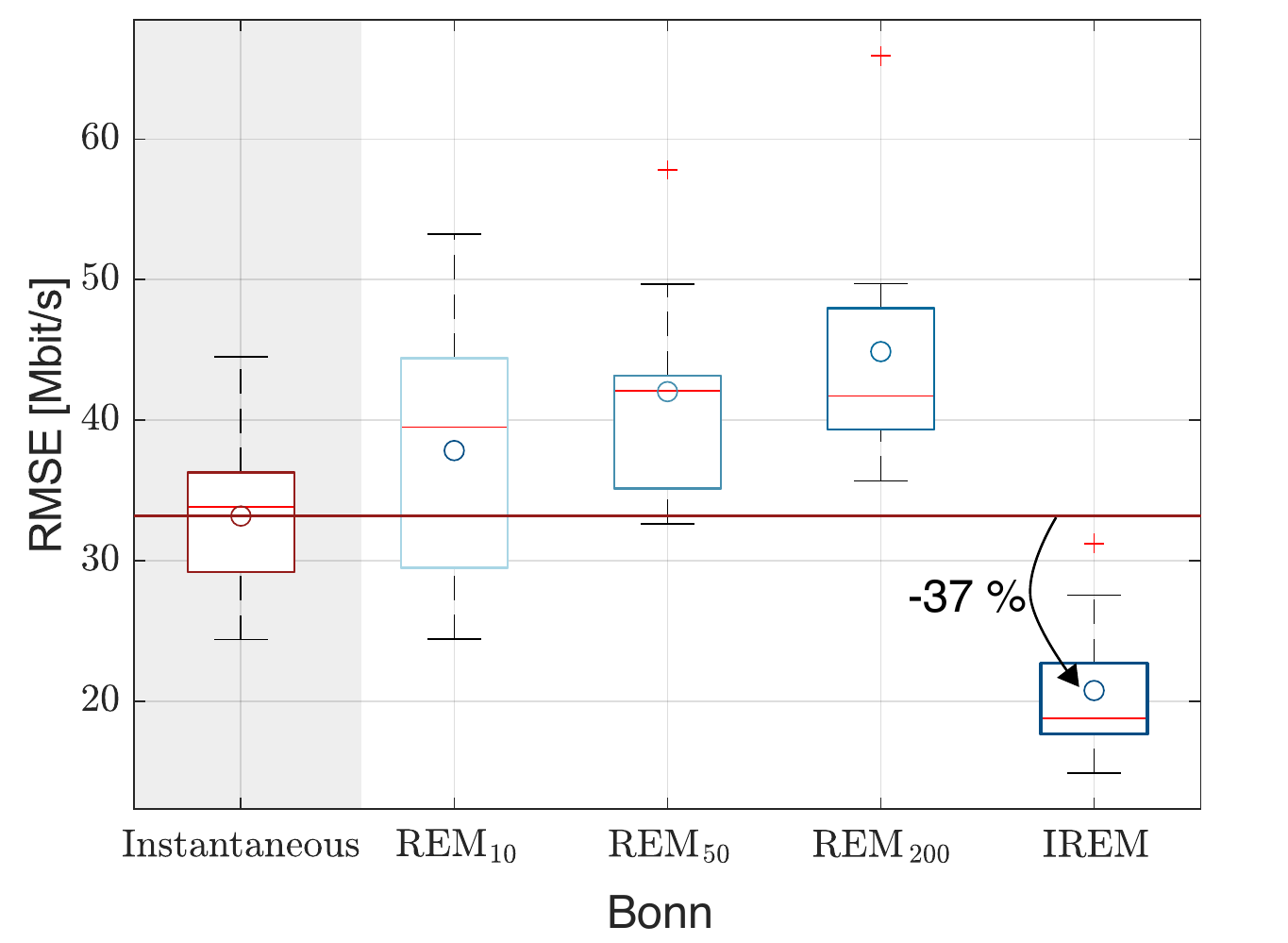}
	\caption{Data rate prediction performance comparison between \acp{REM} with a constant grid size, an optimized \ac{REM} with individually tuned grid sizes for every feature (IREM) and an instantaneous prediction as the baseline. The results are exemplarily evaluated at the Bonn scenario.}
	\label{fig:remgridsizeadaption}
\end{figure}

\textbf{\ac{REM}-based and Combined Prediction}: In \autoref{fig:datarateprediction}, the prediction error of different \ac{REM} configurations is compared to instantaneous predictions based on real-time channel measurements. One \ac{REM} setup considers only the passively measured \acp{RS}. A second \ac{REM} also utilizes high-level and advanced features to predict the \ac{UDP} uplink data rate. Lastly, a combined approach both using real-time measured data and \ac{REM} data is evaluated. For the \acp{REM} a grid size $c$ of \SI{50}{\metre} is used.\\
A \acl{RF} (\acs{RF}) machine learning model consisting of 560 trees and a maximum depth of 40 is trained on each configuration. 10-fold cross-validation is used to prevent overfitting. In addition to that, the learning process is repeated ten times with random initialization to get more reliable results.\\
It can be seen that \ac{REM}-based data rate predictions can keep up with instantaneous data rate predictions. Based on the chosen scenario, \ac{REM}-based predictions can even outperform real-time predictions. If \ac{REM} data is available in addition to real-time \acp{RS} -- Combined Approach --, data rate prediction can be improved in all scenarios. In addition to that, based on the configuration, significant improvements can be reached by using additional parameters in the \ac{REM} (Multidimensional REM). This observation is consistent with the results of the authors of \cite{Palaios/etal/2021a} for instantaneous data rate predictions. However, the improvement is dependent on the scenario and the added features.\\
The cell width $c$ is another design factor affecting the data rate prediction performance of \acp{REM}, as previously mentioned in \autoref{sec:related_work}. To analyze the impact of the cell width on \ac{5G} \ac{NSA} data rate predictions, \acp{REM} with different cell widths are created with a set part of the measurements (training set). On this training data set, an \ac{RF} is trained. Then a data rate prediction on a distinct test set is performed. This approach is repeated ten times for every scenario, to get more reliable results. All available features are used in the \ac{REM} and the same cell width is set for each feature in a \ac{REM}. The results are shown exemplarily for the Bonn scenario.\\
As can be seen in \autoref{fig:remgridsizeadaption}, different cell widths result in varying \acp{RMSE}. A large cell width of \SI{200}{\metre} mostly results in the worst \ac{RMSE} compared to the other \ac{REM} approaches due to the lack of resolution. However, the smallest cell width of \SI{10}{\metre} does not always yield the best \ac{RMSE}: At this cell width, less averaging gains are achieved and more lookup misses occur than at a cell width of \SI{50}{\metre} \cite{Sliwa/etal/2020a}.\\
\textbf{Indicator individually optimized \acp{REM}}: The explained results pose the question, if the prediction accuracy improves, if some features have a higher resolution, while other features are more locally averaged. Some features might have a more pronounced local dependence than others. That is why, an optimized \ac{REM} with individual cell widths for each feature is created --- so-called IREM. This \ac{REM} is optimized using a random search approach for the cell width (10, 20, 50, 75, 100, 200 and \SI{400}{\metre} are considered) with the goal of minimizing the \ac{RMSE} of the data rate prediction. 2000 iterations of the random search are performed to ensure, the search space is sufficiently covered.\\
It can be seen in \autoref{fig:remgridsizeadaption}, that the I\ac{REM} does not only exceeds the prediction performance of \acp{REM} with a set cell size, but also outperforms the instantaneous predictions. A gain of up to \SI{40}{\percent} is reached compared to the instantaneous prediction.\\

\section{Towards Systematic Evaluation of Autonomous Driving}
The evaluation and scalability analysis of \ac{ToD} use cases is a crucial development stage. It needs to preserve a high grade of realism and a high reproducibility of drive tests to underline the significance of the results.
As shown in \autoref{fig:systemigram}, evaluation methods range from analytical models over simulative approaches towards the integration of \ac{HiL} and, finally, real-world experiments.\\
Although \ac{HiL} setups take place in a controlled laboratory environment, scalability is often realized by traffic shapers, and thus, is not suitable to figure the actual characteristics and dynamics of \ac{ToD} scenarios. Nor do cabled setups offer the possibility to carry out a parallel evaluation of the resulting \ac{QoE}. 
On the other hand, real-world trials  can overcome this constraint and provide a complete end-to-end tele-operation but also come at full costs, especially in the case of scalability analyses.\\
\begin{figure}[]  	
	\vspace{0cm}
	\centering		  
	\includegraphics[width=1\columnwidth]{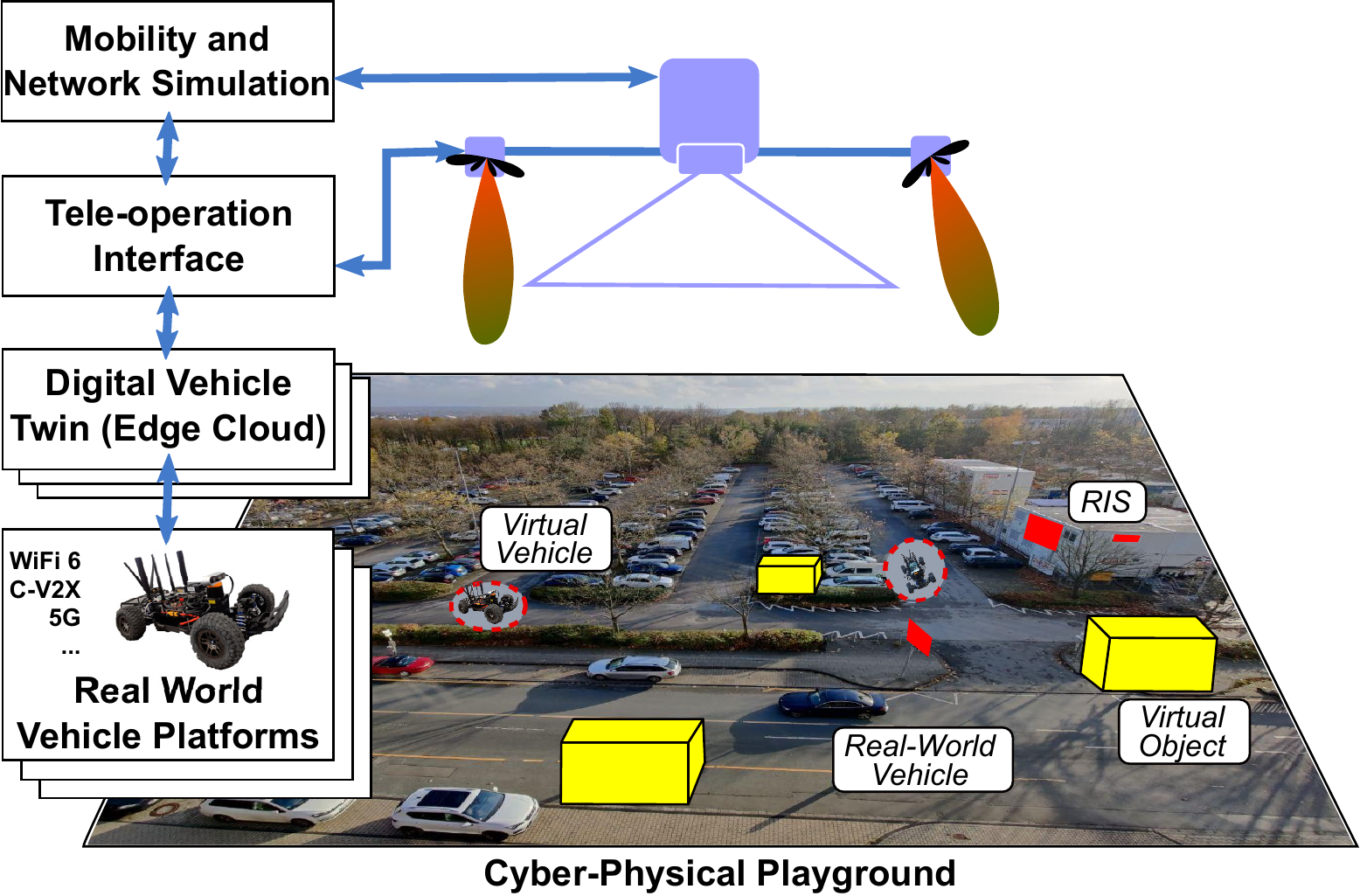}
	\vspace{-0cm}	
	\caption{Cyber-physical playground consisting of real vehicles with respective digital twins in a \ac{ToD} operation configuration.}
	\label{fig:cyberphysicalPlayground}
	\vspace{-0cm}	
\end{figure}

A combination of both approaches in a cyber-physical playground can overcome these issues, as demonstrated in \autoref{fig:cyberphysicalPlayground}. Teleoperated real-world vehicles interact with several real objects and thus experience real behavior and physics. However, all real objects also exist as digital twins. In addition to that, further virtual objects utilizing the \ac{5G} communication technology can be introduced, to which the vehicles need to adapt. These can act as obstacles or restrictions. The reaction of vehicles to these can be observed and improved. As a result, real-world physics with the repeatability of laboratory experiments can be achieved.\\

\begin{figure}[]  	
	\vspace{0cm}
	\centering		  
	\includegraphics[width=.8\columnwidth]{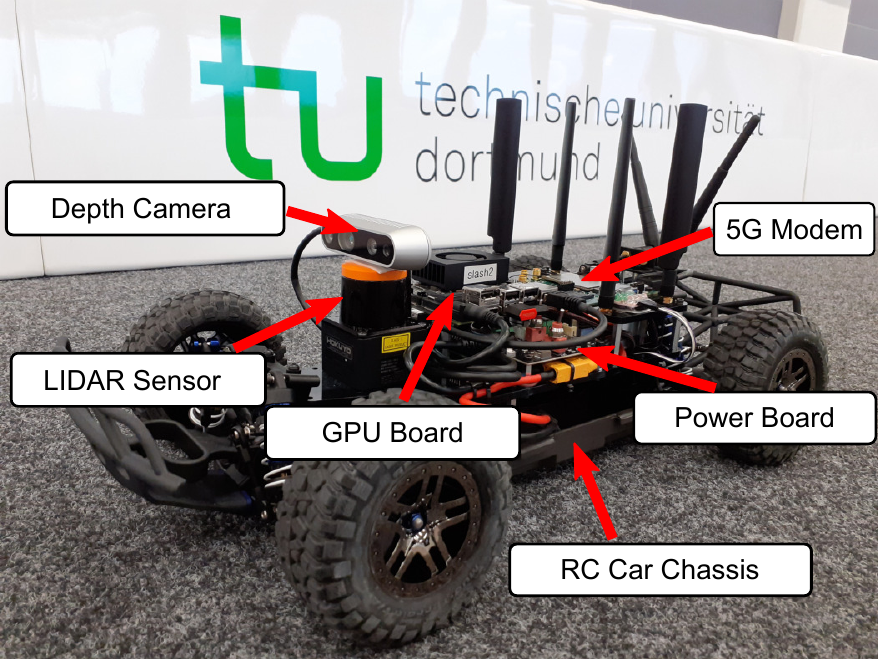}
	\vspace{-0cm}	
	\caption{Scaled vehicular platforms provides a cost-effective opportunity for scalability analyses in controlled environments to evaluate novel approaches.}
	\label{fig:f1tenth}
	\vspace{-0cm}	
\end{figure}

To improve scalability, lower the cost, but maintain the end-to-end aspect of the teleoperator, this setup can be sized down to be used with scaled vehicular platforms. The used vehicle is built in alignment with the \textit{F1/10} project \cite{F1Tenth/2021a} and consists of the ground chassis of a consumer radio-controlled car, which is further equipped with communication equipment, sensors, and computation entities.\\
A detailed picture is shown in \autoref{fig:f1tenth}. Besides the chassis and original drive train, a motor controller is equipped to enable smooth steering of the brush-less motor and servo actuators through networked tele-operation.\\
Further, there are two main computation units. An \textit{NVIDIA Nano NX} adds powerful graphical processing capability. Although it was originally installed for autonomous driving use cases, it can also be utilized to compute complex machine learning algorithms locally. A parallel \textit{Raspberry Pi} orchestrates the \ac{5G} connectivity over \textit{Quectel RM500Q} modules and is, in the long-term, to be extended with other communication technologies to support multi-\ac{RAT} approaches.
A high-resolution depth camera in the front of the car delivers rich visual material. In extension to this, the \ac{LIDAR} senses the environmental information to enhance the teleoperator's perception.\\
These F1/10 based scaled vehicles can be used to enable large-scale tele-operated driving at indoor and outdoor test sites. Before non-scaled vehicles are ready for testing, complex maneuvers can be driven using the real-world radio channel. This way, the development of \ac{ToD} platforms can be accelerated in an effective and relatively cost-effective way. In addition, trials with a scaled vehicular platform do not pose a potential danger for test personnel that would need to be in a tele-operated vehicle for emergency halting.\\

\begin{figure}[]  	
	\vspace{0cm}
	\centering		  
	\includegraphics[width=1\columnwidth]{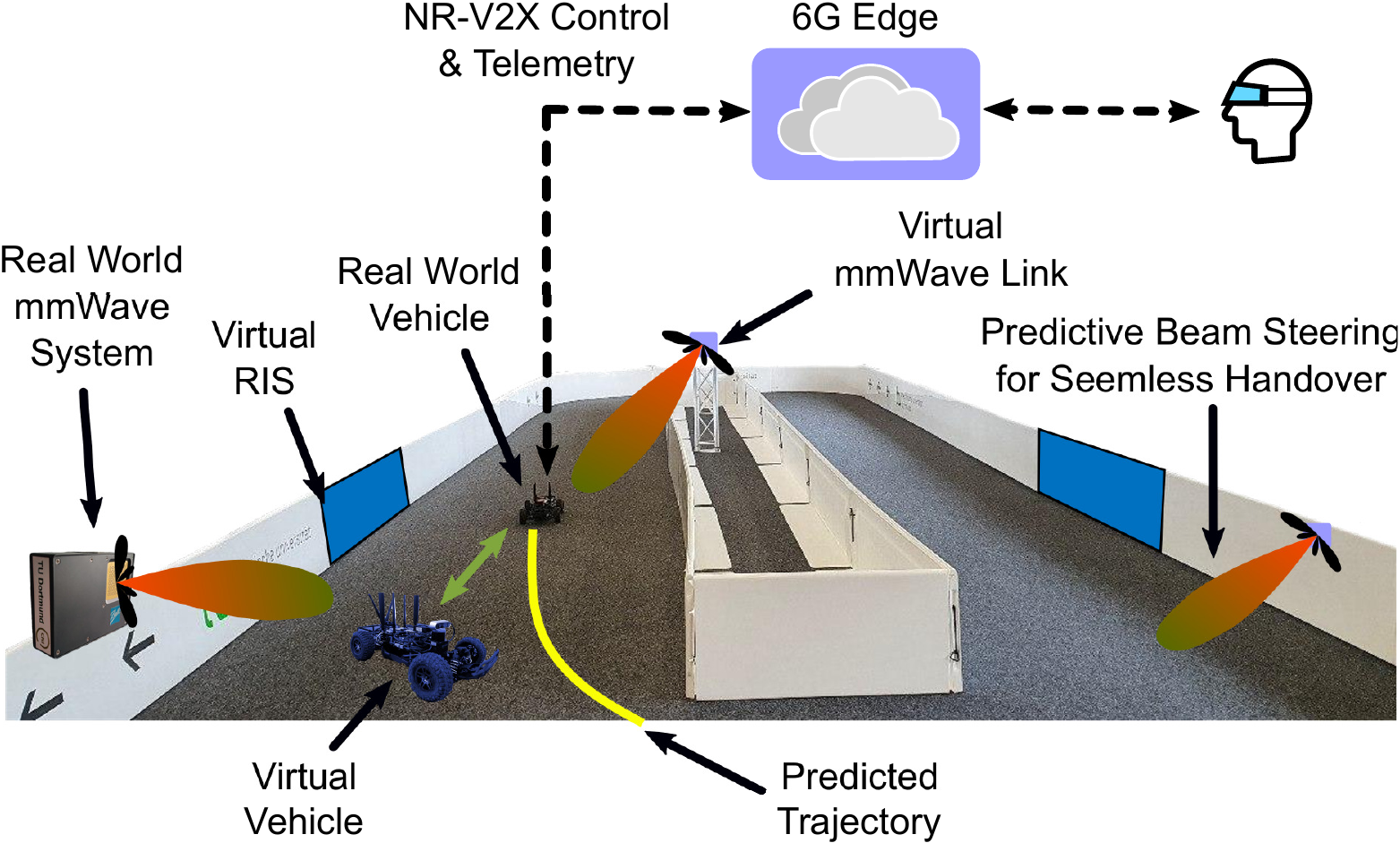}
	\vspace{-0cm}	
	\caption{A cyber-physical playground can be used for tele-operated driving evaluation with scaled vehicular platforms connected via a \ac{6G} network and \ac{V2X} connectivity, including trajectory prediction and virtual \aclp{RIS} (\acsp{RIS}).}
	\label{fig:f110Playground}
	\vspace{-0cm}	
\end{figure}
The described evaluation setup has the potential to be used beyond \ac{5G}, for example, to test \acp{RIS}, which are a key enabler of future \ac{6G} networks. A playground for these F1/10 vehicles based on \ac{6G} \ac{mmWave} and \ac{V2X} technology is shown in \autoref{fig:f110Playground}. The trajectories of the real-world vehicles are prognosticated to enable predictive beam steering for seamless handovers. To test the system, virtual \acp{RIS} can be introduced, while a teleoperator controls the F1/10 vehicle, driving through the environment consisting of real and virtual objects and vehicles. That way, teleoperation with \ac{6G} \ac{mmWave} technology can be tested in different challenging virtual environments.

\section{Conclusion}

For tele-operated driving, multiple individual systems need to interact with each other. Various challenges from several research directions need to be overcome to enable \ac{ToD} smoothly and safely. One part of this system is the mobile network part.\\
Different data rate requirements are specified in related work. The current \ac{5G} \ac{NSA} mobile network can not always fulfill these. That is why in advance predictive \ac{QoS} is a key enabler of \ac{ToD}. In the case of a insufficient data rate, alternative routes can be chosen, or the velocity adapted. One method to predict in advance end-to-end data rates are so-called \acp{REM}. The results of \ac{REM}-based predictions are comparable to instantaneous predictions. However, with the help of individually optimized \acp{REM}, which tune the cell size of each layer individually, the prediction accuracy can be further improved. In the future, current \ac{UE}-based predictions might be further improved by network slicing and network-side-based data rate predictions \cite{Palaios/etal/2021a}. \\
Since \ac{ToD} is a complex system, the evaluation of tele-operated driving systems is a challenging task. In this paper, an evaluation approach for \ac{ToD} and autonomous driving based on cyber-physical playgrounds is given. With the help of real and virtual vehicles and obstacles, real-world physical interaction similar to the final product can be achieved with the repeatability of laboratory trials. By sizing down this approach to scaled vehicular platforms, which are safer to handle, further cost reductions can be achieved.

\ifdoubleblind

\else

	\section*{Acknowledgment}
	
	\footnotesize
		%
	%
	This work has received funding by the German Federal Ministry of Education and Research
	(BMBF) in the course of the \emph{6GEM} research hub under grant number 16KISK038 
	%
	%
	and by the German Research Foundation (DFG) within the \emph{Collaborative Research Center SFB 876} ``Providing Information by Resource-Constrained Analysis'', projects A4 and B4,
	%
	%
	as well as by the Ministry of Economic Affairs, Innovation, Digitalization and Energy of the state of North Rhine--Westphalia (MWIDE NRW) in the course of the \emph{Competence Center 5G.NRW} under grant number 005--01903--0047, the \emph{Plan \& Play} under grant number 005-2008-0047
	%
	%
	and has received funding by the Federal Ministry of Transport and Digital Infrastructure (BMVI) in the context of the project \emph{Virtual integration of decentralized charging infrastructure in cab stands} under the funding reference 16DKVM006B.

\fi

\balance
\ifacm
\else
	\bibliographystyle{IEEEtran}
	\bibliography{Bibliography}
\fi

\end{document}